\documentstyle[12pt,qqa4lart]{article}
\input tcilatex

\QQQ{Language}{
American English
}

\begin{document}

\author{Lu-Ming Duan and Guang-Can Guo\thanks{
E-mail:gcguo@sunlx06.nsc.ustc.edu.cn} \\
Physics Department and Nonlinear Science Center, University\\
of Science and Technology of China, Hefei,230026 P.R.China}
\title{Alternative approach to electromagnetic field quantization in nonlinear and
inhomogeneous media}
\date{}
\maketitle

\begin{abstract}
\baselineskip 24pt A simple approach is proposed for the quantization of the
electromagnetic field in nonlinear and inhomogeneous media. Given the
dielectric function and nonlinear susceptibilities, the Hamiltonian of the
electromagnetic field is determined completely by this quantization method.
From Heisenberg's equations we derive Maxwell's equations for the field
operators. When the nonlinearity goes to zero, this quantization method
returns to the generalized canonical quantization procedure for linear
inhomogeneous media [Phys. Rev. A, 43, 467, 1991]. The explicit Hamiltonians
for the second-order and third-order nonlinear quasi-steady-state processes
are obtained based on this quantization procedure.

\ 

\end{abstract}

\newpage\baselineskip 24pt

\section{Introduction}

Early quantization of the electromagnetic field is performed in empty
cavities or in infinite free space [1]. However, with the growth of interest
in quantum optical phenomena taking place inside material media, several
approaches have been proposed for quantization of the electromagnetic field
in nonlinear, inhomogeneous, or dispersive media [2-21]. Early attempts
towards quantization of the nonlinear media, while incorporating the known
linear theory, did not fully reproduce the nonlinear field equations [4]. An
innovative treatment was first proposed by Hillery and Mlodinow who
successfully quantized a nonlinear medium by introducing the dual potential
[6]. Later, Drummond extended the Hillery-Mlodinow procedure to dispersive
media [9]. There are also other approaches in this direction. Glauber and
Lewenstein generalized the canonical quantization method by modifying the
gauge condition to deal with the inhomogeneous linear media [11]. Abram and
Cohen, following the canonical quantization procedure, presented a quantum
formulation for light propagation in nonlinear effective media [12]. And
recently, Santos and Loudon gave an alternative approach to the quantization
of the electromagnetic field in linear one-dimensional dispersive media
[19]. Developments towards the absorbing dielectrics also appeared [20,21].

In this paper, we propose a relatively simpler approach to the
electromagnetic field quantization in nonlinear and inhomogeneous media. The
procedure follows Ref. [11] in using the material independence of the
commutation relations for the fields $\overrightarrow{D}$ and $%
\overrightarrow{B}$, pointed out by Born and Infeld [2], as a starting point
in the quantization. We extend this to nonlinear media in which $%
\overrightarrow{D}$ and $\overrightarrow{B}$ can be expressed as isochronous
functionals of the fields $\overrightarrow{E}$ and $\overrightarrow{H}$. $%
\overrightarrow{D}$ and $\overrightarrow{B}$ are expanded into the mode
functions. Furthermore, we explicitly derive Maxwell's equations for the
field operators from Heisenberg's equations. This procedure is applied to
the quantization of the second-order and third-order nonlinear
quasi-steady-state processes and we obtain the explicit Hamiltonians. Though
these Hamiltonians are already in wide use in quantum optics, their
derivations are mainly based on the early quantization procedure by Shen
[4,22,23] and known by now to be inconsistent with the nonlinear field
equations [9]. So here we give a justification of these Hamiltonians.

The arrangement of the paper is as follows. The quantization procedure is
proposed in Sec.2. Given dielectric tensor and nonlinear susceptibilities,
this quantization procedure completely determines the Hamiltonian of the
electromagnetic field, which is expressed by annihilation and creation
operators. Then we derive Maxwell's equations from Heisenberg's equations
for the field operators. In Sec.3 we show this quantization procedure
returns to the generalized canonical quantization method in Ref. [11] when
the medium is linear. The explicit Hamiltonians of the second-order and
third-order nonlinear quasi-steady-state processes are obtained in Sec. 4 by
application of this quantization procedure.

\section{Quantization in the presence of nonlinear media}

We consider the electromagnetic field in nonlinear media, which may be
inhomogeneous. The source-free Maxwell equations in matter take the forms
[24] 
\begin{equation}
\label{1}\nabla \cdot \overrightarrow{D}=0,
\end{equation}
\begin{equation}
\label{2}\nabla \cdot \overrightarrow{B}=0,
\end{equation}
\begin{equation}
\label{3}\frac 1c\frac{\partial \overrightarrow{D}}{\partial t}=\nabla
\times \overrightarrow{H},
\end{equation}
\begin{equation}
\label{4}\frac 1c\frac{\partial \overrightarrow{B}}{\partial t}=-\nabla
\times \overrightarrow{E}.
\end{equation}
In nonlinear media, $\overrightarrow{D}\left( t\right) $ and $%
\overrightarrow{B}\left( t\right) $ are complicated nonlinear functionals of 
$\overrightarrow{E}\left( t\right) $ and $\overrightarrow{H}\left( t\right) $%
. From the Maxwell equations, the energy density $U$ of the electromagnetic
field in nonlinear media is determined by 
\begin{equation}
\label{5}dU\left( \overrightarrow{r},t\right) =\overrightarrow{E}\left( 
\overrightarrow{r},t\right) \cdot d\overrightarrow{D}\left( \overrightarrow{r%
},t\right) +\overrightarrow{H}\left( \overrightarrow{r},t\right) \cdot d%
\overrightarrow{B}\left( \overrightarrow{r},t\right) .
\end{equation}
The Hamiltonian (or the energy) is 
\begin{equation}
\label{6}\widetilde{H}=\int d^3\overrightarrow{r}U\left( \overrightarrow{r}%
,t\right) .
\end{equation}
For the electromagnetic field in linear media, the canonical quantization
method is generally used. The vector potential $\overrightarrow{A}$ is
chosen as the general coordinate and the Columb gauge $\nabla \cdot 
\overrightarrow{A}=0$ is often used. $\overrightarrow{A}$ and its conjugate
momentum can be expanded into a set of transverse complete spatial functions
and the expansion coefficients are expressed by annihilation and creation
operators. Then substituting the expansions of $\overrightarrow{A}$ and its
conjugate momentum into the Hamiltonian, one achieves quantization of the
electromagnetic field in linear media. However, for the nonlinear media, the
canonical quantization becomes much more involved. $\overrightarrow{A}$ and $%
\overrightarrow{E}$ were chosen as the canonical variables in the early
treatments, which did not incorporate Eq.(1). In fact, no rigorous approach
had been proposed for nonlinear media until Hillery and Mlodinow introduced
the dual potential and then followed the canonical quantization procedure.
Here, inspired by the result in Ref.[1] that the fields $\overrightarrow{D}$
and $\overrightarrow{B}$ have medium-independent commutation relations in
inhomogeneous linear media, we choose the fields $\overrightarrow{D}$ and $%
\overrightarrow{B},$ rather than $\overrightarrow{E}$ or $\overrightarrow{A},
$ as the starting point of the electromagnetic field quantization. This
choice is also consistent with the results in Ref. [6] [9] and [12], where
the field $\overrightarrow{D}$ was found to be the canonical momentum.
Starting from the mode expansions of the fields $\overrightarrow{D}$ and $%
\overrightarrow{B}$, we can present a concise formulation of the
quantization and a clear derivation of Maxwell's equations for the field
operators.

From Equations (1) and (2), the fields $\overrightarrow{D}$ and $%
\overrightarrow{B}$ can be expanded into a set of transverse complete
spatial functions $\left\{ \overrightarrow{f}_{\overrightarrow{k}\mu }\left( 
\overrightarrow{r}\right) \right\} $ and $\left\{ \nabla \times 
\overrightarrow{f}_{\overrightarrow{k}\mu }\left( \overrightarrow{r}\right)
\right\} ,$ respectively, 
\begin{equation}
\label{7}\overrightarrow{D}\left( \overrightarrow{r},t\right) =-\stackunder{%
\overrightarrow{k}\mu }{\sum }P_{\overrightarrow{k}\mu }\left( t\right) 
\overrightarrow{f}_{\overrightarrow{k}\mu }^{*}\left( \overrightarrow{r}%
\right) , 
\end{equation}
\begin{equation}
\label{8}\overrightarrow{B}\left( \overrightarrow{r},t\right) =c\stackunder{%
\overrightarrow{k}\mu }{\sum }Q_{\overrightarrow{k}\mu }\left( t\right)
\nabla \times \overrightarrow{f}_{\overrightarrow{k}\mu }\left( 
\overrightarrow{r}\right) . 
\end{equation}
The expansion functions and coefficients satisfy Hermitian conditions

\begin{equation}
\label{9}\overrightarrow{f}_{\overrightarrow{k}\mu }^{*}=\overrightarrow{f}%
_{-\overrightarrow{k}\mu }, 
\end{equation}
\begin{equation}
\label{10}Q_{\overrightarrow{k}\mu }^{+}=Q_{-\overrightarrow{k}\mu }, 
\end{equation}
\begin{equation}
\label{11}P_{\overrightarrow{k}\mu }^{+}=P_{-\overrightarrow{k}\mu }. 
\end{equation}
In addition, the functions $\overrightarrow{f}_{\overrightarrow{k}\mu
}\left( \overrightarrow{r}\right) $ satisfy transversality, orthonormality
and completeness conditions 
\begin{equation}
\label{12}\nabla \cdot \overrightarrow{f}_{\overrightarrow{k}\mu }=0, 
\end{equation}
\begin{equation}
\label{13}\int d^3\overrightarrow{r}\overrightarrow{f}_{\overrightarrow{k}%
\mu i}^{*}\left( \overrightarrow{r}\right) \overrightarrow{f}_{%
\overrightarrow{k}^{^{\prime }}\mu ^{^{\prime }}j}\left( \overrightarrow{r}%
\right) =\delta _{\overrightarrow{k}\overrightarrow{k}^{^{\prime }}}\delta
_{\mu \mu ^{^{\prime }}}\delta _{ij}, 
\end{equation}
\begin{equation}
\label{14}\stackunder{\overrightarrow{k}\mu }{\sum }\overrightarrow{f}_{%
\overrightarrow{k}\mu i}^{*}\left( \overrightarrow{r}\right) \overrightarrow{%
f}_{\overrightarrow{k}\mu j}\left( \overrightarrow{r}^{^{\prime }}\right)
=\delta _{ij}^T\left( \overrightarrow{r}-\overrightarrow{r}^{^{\prime
}}\right) , 
\end{equation}
where the transverse $\delta -$function $\delta _T$ is defined as 
\begin{equation}
\label{15}\delta _{ij}^T\left( \overrightarrow{r}\right) =\frac 1{\left(
2\pi \right) ^3}\int d^3\overrightarrow{k}\left( \delta _{ij}-\frac{k_ik_j}{%
\left| \overrightarrow{k}\right| ^2}\right) e^{i\overrightarrow{k}\cdot 
\overrightarrow{r}}. 
\end{equation}
The transversality condition (12) makes the completeness equation of $%
\overrightarrow{f}_{\overrightarrow{k}\mu }\left( \overrightarrow{r}\right) $
take the form of (14). In free space, the plane wave is chosen as the
expansion function 
\begin{equation}
\label{16}\overrightarrow{f}_{\overrightarrow{k}\mu }\left( \overrightarrow{r%
}\right) =\frac 1{\left( 2\pi \right) ^{\frac 32}}\overrightarrow{e}_{%
\overrightarrow{k}\mu }e^{i\overrightarrow{k}\cdot \overrightarrow{r}}, 
\end{equation}
where the unit vectors $\overrightarrow{e}_{\overrightarrow{k}\mu }$ $\left(
\mu =1,2\right) $ satisfy 
\begin{equation}
\label{17}\overrightarrow{k}\cdot \overrightarrow{e}_{\overrightarrow{k}\mu
}=0. 
\end{equation}

The expansions (7) and (8) have the same forms as those in linear media. We
further suppose that the expansion coefficient operators satisfy the same
commutation relations. So 
\begin{equation}
\label{18}\left[ Q_{\overrightarrow{k}\mu }\left( t\right) ,P_{%
\overrightarrow{k}^{^{\prime }}\mu ^{^{\prime }}}\left( t\right) \right]
=i\hbar \delta _{\overrightarrow{k}\overrightarrow{k}^{^{\prime }}}\delta
_{\mu \mu ^{^{\prime }}}. 
\end{equation}
The fields $\overrightarrow{E}\left( \overrightarrow{r},t\right) $ and $%
\overrightarrow{H}\left( \overrightarrow{r},t\right) $ can be expressed by $%
\overrightarrow{D}\left( \overrightarrow{r},t\right) $ and $\overrightarrow{B%
}\left( \overrightarrow{r},t\right) $ from the nonlinear functional
relations between them. From (5) and (6) the Hamiltonian of the
electromagnetic field becomes a nonlinear functional of $\overrightarrow{D}%
\left( \overrightarrow{r},t\right) $ and $\overrightarrow{B}\left( 
\overrightarrow{r},t\right) $. After substituting the expansions (7) and (8)
into it , we get the Hamiltonian, which is expressed by annihilation and
creation operators. Given the functional relations between $\overrightarrow{E%
}\left( \overrightarrow{r},t\right) $,$\overrightarrow{H}\left( 
\overrightarrow{r},t\right) $ and $\overrightarrow{D}\left( \overrightarrow{r%
},t\right) $,$\overrightarrow{B}\left( \overrightarrow{r},t\right) $, the
Hamiltonian form is completely determined by this quantization procedure.

Now we show reasonableness of the quantization method. Comparing (7),(8) and
(18) with the corresponding equations in Ref.[1], we know that when the
nonlinearity goes to zero the above procedure returns to the generalized
canonical quantization method. Furthermore, this quantization gives the
correct Maxwell equations. In the following we derive Maxwell's equations
for the field operators from Heisenberg's equations.

From the transversality of the expansion functions $\overrightarrow{f}_{%
\overrightarrow{k}\mu },$ the first two Maxwell equations (1) and (2) are
obviously satisfied. Equations (7) (8) and (14) (18) give the commutator of
the field operators $\overrightarrow{D}$ and $\overrightarrow{B}$%
\begin{equation}
\label{19}\left[ D_i\left( \overrightarrow{r},t\right) ,B_j\left( 
\overrightarrow{r}^{^{\prime }},t\right) \right] =i\hbar c\left( \nabla
\times \delta \right) _{ij}\left( \overrightarrow{r}-\overrightarrow{r}%
^{^{\prime }}\right) . 
\end{equation}
In the derivation, the following relations are used. 
\begin{equation}
\label{20}
\begin{array}{c}
\left( \nabla \times \delta ^T\right) _{ij}\left( 
\overrightarrow{r}\right) =\stackunder{mn}{\sum }\varepsilon _{imn}\partial
_m\frac 1{\left( 2\pi \right) ^3}\int d^3\overrightarrow{k}\left( \delta
_{nj}-\frac{k_nk_j}{\left| \overrightarrow{k}\right| ^2}\right) e^{i%
\overrightarrow{k}\cdot \overrightarrow{r}} \\ = 
\stackunder{mn}{\sum }\varepsilon _{imn}\partial _m\frac 1{\left( 2\pi
\right) ^3}\int d^3\overrightarrow{k}\delta _{nj}e^{i\overrightarrow{k}\cdot 
\overrightarrow{r}} \\ =\left( \nabla \times \delta \right) _{ij}\left( 
\overrightarrow{r}\right) 
\end{array}
\end{equation}
and 
\begin{equation}
\label{21}
\begin{array}{c}
\left( \nabla ^{^{\prime }}\times \delta \right) _{ji}\left( 
\overrightarrow{r}-\overrightarrow{r}^{^{\prime }}\right) =-\left( \nabla
\times \delta \right) _{ji}\left( \overrightarrow{r}-\overrightarrow{r}%
^{^{\prime }}\right) \\ =- 
\stackunder{mn}{\sum }\varepsilon _{jmn}\partial _m\delta _{ni}\left( 
\overrightarrow{r}-\overrightarrow{r}^{^{\prime }}\right) \\ =- 
\stackunder{m}{\sum }\varepsilon _{jmi}\partial _m\delta \left( 
\overrightarrow{r}-\overrightarrow{r}^{^{\prime }}\right) \\ = 
\stackunder{m}{\sum }\varepsilon _{imj}\partial _m\delta \left( 
\overrightarrow{r}-\overrightarrow{r}^{^{\prime }}\right) \\ =\left( \nabla
\times \delta \right) _{ij}\left( \overrightarrow{r}-\overrightarrow{r}%
^{^{\prime }}\right) \text{ }. 
\end{array}
\end{equation}
The commutator (19) has been given in Ref. [12] in one-dimensional form.
Here we extend it to the general case. From the commutator (19), the
commutator between $\overrightarrow{D}$ or $\overrightarrow{B}$ and an
arbitrary functional $F\left( \overrightarrow{D},\overrightarrow{B}\right) $
of $\overrightarrow{D}$ and $\overrightarrow{B},$ which may be nonlinear,
can be expressed by functional derivation as follows: 
\begin{equation}
\label{22}\left[ D_i\left( \overrightarrow{r},t\right) ,F\right] =i\hbar c%
\stackunder{mn}{\sum }\varepsilon _{imn}\partial _m\frac \delta {\delta
B_n\left( \overrightarrow{r},t\right) }F, 
\end{equation}
\begin{equation}
\label{23}\left[ B_i\left( \overrightarrow{r},t\right) ,F\right] =-i\hbar c%
\stackunder{mn}{\sum }\varepsilon _{imn}\partial _m\frac \delta {\delta
D_n\left( \overrightarrow{r},t\right) }F. 
\end{equation}
From these two equations and (5),(6), the Heisenberg equations of the field
operators $\overrightarrow{D}$ and $\overrightarrow{B}$ take the forms 
\begin{equation}
\label{24}
\begin{array}{c}
\frac 1c 
\frac{\partial D_i}{\partial t}=\frac 1{i\hbar c}\left[ D_i\left( 
\overrightarrow{r},t\right) ,\widetilde{H}\right] \\  \\ 
= 
\stackunder{mn}{\sum }\varepsilon _{imn}\partial _m\frac \delta {\delta
B_n\left( \overrightarrow{r},t\right) }\widetilde{H} \\  \\ 
= 
\stackunder{mn}{\sum }\varepsilon _{imn}\partial _m\frac{\partial U\left( 
\overrightarrow{r},t\right) }{\partial B_n\left( \overrightarrow{r},t\right) 
} \\  \\ 
= 
\stackunder{mn}{\sum }\stackunder{j}{\sum }\varepsilon _{imn}\partial _mH_j%
\frac{\partial B_j}{\partial B_n} \\  \\ 
=\left( \nabla \times \overrightarrow{H}\right) _i\text{ .} 
\end{array}
\end{equation}
Similarly 
\begin{equation}
\label{25}\frac 1c\frac{\partial B_i}{\partial t}=-\left( \nabla \times 
\overrightarrow{E}\right) _i\text{ .} 
\end{equation}
So we have clearly derived Maxwell's equations for the field operators from
Heisenberg's equations. The isochronous commutators (22),(23) play an
essential role in the derivation. The derivation holds for linear or
nonlinear media. However, for dispersive media, nonlocal relations in time
between the Hamiltonian $\widetilde{H}$ and the fields $\overrightarrow{D},%
\overrightarrow{B}$ arise and the isochronous commutators (22),(23) cannot
be applied in this case. Here we meet the long standing difficulty in
quantum optics in quantizing nonlinear and dispersive dielectrics.

\section{Quantization of the electromagnetic field in linear inhomogeneous
media}

In this section we use the above method to quantize the electromagnetic
field in linear inhomogeneous media. The medium is characterized by 
\begin{equation}
\label{26}D_i\left( \overrightarrow{r},t\right) =\stackunder{j}{\sum }%
\varepsilon _{ij}\left( \overrightarrow{r}\right) E_j\left( \overrightarrow{r%
},t\right) , 
\end{equation}
\begin{equation}
\label{27}\overrightarrow{B}\left( \overrightarrow{r},t\right) =%
\overrightarrow{H}\left( \overrightarrow{r},t\right) . 
\end{equation}
The Hamiltonian (or the energy) is 
\begin{equation}
\label{28}\widetilde{H}=\frac 12\int d^3\overrightarrow{r}\left( 
\overrightarrow{B}^2+\stackunder{ij}{\sum }\varepsilon
_{ij}^{-1}D_iD_j\right) . 
\end{equation}

In free space the expansion function $\overrightarrow{f}_{\overrightarrow{k}%
\mu }$ is expressed by Eq.(16). Substituting the expansions (7) and (8) into
Eq.(28) and noting 
\begin{equation}
\label{29}Q_{\overrightarrow{k}\mu }\left( t\right) =\left( \frac \hbar
{2\omega _{\overrightarrow{k}\mu }}\right) ^{\frac 12}\left( a_{%
\overrightarrow{k}\mu }+a_{-\overrightarrow{k}\mu }^{+}\right) , 
\end{equation}
\begin{equation}
\label{30}P_{\overrightarrow{k}\mu }\left( t\right) =i\left( \frac{\hbar
\omega _{\overrightarrow{k}\mu }}2\right) ^{\frac 12}\left( a_{%
\overrightarrow{k}\mu }^{+}-a_{-\overrightarrow{k}\mu }\right) , 
\end{equation}
where $\omega _{\overrightarrow{k}\mu }=\left| \overrightarrow{k}\right| c$
, we get the Hamiltonian expressed by annihilation and creation operators. 
\begin{equation}
\label{31}
\begin{array}{c}
\widetilde{H}=\stackunder{_{\overrightarrow{k}\mu }}{\sum }\hbar \omega _{%
\overrightarrow{k}\mu }a_{\overrightarrow{k}\mu }^{+}a_{\overrightarrow{k}%
\mu }+\frac \hbar 4\stackunder{_{\overrightarrow{k}\mu }}{\sum }\stackunder{%
_{\overrightarrow{k}^{^{\prime }}\mu ^{^{\prime }}}}{\sum }\sqrt{\omega _{%
\overrightarrow{k}\mu }\omega _{\overrightarrow{k}^{^{\prime }}\mu
^{^{\prime }}}} \\ \times \left[ V_{\mu \mu ^{^{\prime }}}^{*}\left( 
\overrightarrow{k},\overrightarrow{k}^{^{\prime }}\right) a_{\overrightarrow{%
k}\mu }^{+}a_{\overrightarrow{k}^{^{\prime }}\mu ^{^{\prime }}}^{+}-V_{\mu
\mu ^{^{\prime }}}^{*}\left( \overrightarrow{k},-\overrightarrow{k}%
^{^{\prime }}\right) a_{\overrightarrow{k}\mu }^{+}a_{\overrightarrow{k}%
^{^{\prime }}\mu ^{^{\prime }}}+h.c.\right] , 
\end{array}
\end{equation}
where $V_{\mu \mu ^{^{\prime }}}$ is defined by 
\begin{equation}
\label{32}V_{\mu \mu ^{^{\prime }}}\left( \overrightarrow{k},\overrightarrow{%
k}^{^{\prime }}\right) =\frac 1{\left( 2\pi \right) ^3}\int d^3%
\overrightarrow{r}\overrightarrow{e}_{\overrightarrow{k}\mu }\cdot \left(
1-\epsilon ^{-1}\right) \cdot \overrightarrow{e}_{\overrightarrow{k}%
^{^{\prime }}\mu ^{^{\prime }}}e^{i\left( \overrightarrow{k}+\overrightarrow{%
k}^{^{\prime }}\right) \cdot \overrightarrow{r}} 
\end{equation}
If the second-order tensor $\varepsilon ^{-1}$ is a scalar, $\overrightarrow{%
e}_{\overrightarrow{k}\mu }\cdot \overrightarrow{e}_{\overrightarrow{k}%
^{^{\prime }}\mu ^{^{\prime }}}$ in Eq.(32) can be put out of the
integration. This case had been discussed in detail in Ref.[11] by the
generalized canonical quantization method. Here we get the same results.

\section{Quantization of the quasi-steady-state optical field in nonlinear
media}

In this section we consider quantization of the quasi-steady-state optical
field in nonlinear media. The second-order or third-order nonlinear process
is most important [25,23]. First we apply the quantization method to the
parametric process. The optical field is composed of three
quasi-monochromatic fields with central frequency $\omega _1,\omega
_2,\omega _3$, respectively, and $\omega _3=\omega _1+\omega _2$, i.e., 
\begin{equation}
\label{33}\overrightarrow{E}\left( t\right) =\stackrel{3}{\stackunder{i=1}{%
\sum }}\overrightarrow{E}^{\left( i\right) }\left( t\right) e^{-i\omega
_it}+h.c., 
\end{equation}
where $\overrightarrow{E}^{\left( i\right) }\left( t\right) $ is slowly
varying amplitude. Under the quasi-steady-state approximation, the term $%
\overrightarrow{E}^{\left( i\right) }\left( t\right) e^{-i\omega _it}$ in
Eq.(33) can be viewed as a monochromatic field with frequency $\omega _i$,
i.e., the dispersion of the optical field in the medium is negligible.
Suppose the refractive index is independent of orientation of space. Then
the field $\overrightarrow{D}^{\left( i\right) }\left( t\right) $ can be
expressed by $\overrightarrow{E}^{\left( i\right) }\left( t\right) $. For
example, 
\begin{equation}
\label{34}\overrightarrow{D}^{\left( 3\right) }\left( t\right) =n^2\left(
\omega _3\right) \overrightarrow{E}^{\left( 3\right) }\left( t\right) +\chi
^{\left( 2\right) }\left( \omega _3=\omega _1+\omega _2\right) :%
\overrightarrow{E}^{\left( 1\right) }\left( t\right) \overrightarrow{E}%
^{\left( 2\right) }\left( t\right) , 
\end{equation}
where $n^2\left( \omega _3\right) $ is introduced phenomenologically to show
the component of the optical field $\overrightarrow{E}^{\left( 3\right)
}\left( t\right) $ has been viewed as a monochromatic field with frequency $%
\omega _3$. $\overrightarrow{D}^{\left( 2\right) }\left( t\right) $ and $%
\overrightarrow{D}^{\left( 1\right) }\left( t\right) $ have similar
expressions. From these equations, $\overrightarrow{E}^{\left( i\right)
}\left( t\right) $ $\left( i=1,2,3\right) $ may be expressed by $%
\overrightarrow{D}^{\left( i\right) }\left( t\right) $ as 
\begin{equation}
\label{35}\overrightarrow{E}^{\left( 3\right) }\left( t\right) =\frac{%
\overrightarrow{D}^{\left( 3\right) }\left( t\right) }{n^2\left( \omega
_3\right) }-\gamma ^{\left( 2\right) }\left( \omega _3=\omega _1+\omega
_2\right) :\overrightarrow{D}^{\left( 1\right) }\left( t\right) 
\overrightarrow{D}^{\left( 2\right) }\left( t\right) ,\text{ et.al.,} 
\end{equation}
where 
\begin{equation}
\label{36}\gamma ^{\left( 2\right) }\left( \omega _3=\omega _1+\omega
_2\right) =\frac 1{n^2\left( \omega _1\right) n^2\left( \omega _2\right)
n^2\left( \omega _3\right) }\chi ^{\left( 2\right) }\left( \omega _3=\omega
_1+\omega _2\right) . 
\end{equation}
In the derivation of Eq.(35) the approximation that the nonlinear terms are
much smaller than the linear terms is used. Substituting (35) into (5) and
(6) and making the rotating wave approximation, we get the Hamiltonian of
the electromagnetic field, which is expressed by $\overrightarrow{D}$ and $%
\overrightarrow{B}$%
\begin{equation}
\label{37}
\begin{array}{c}
\widetilde{H}=\frac 12\int d^3\overrightarrow{r}\left\{ \stackrel{3}{%
\stackunder{i=1}{\sum }}\left[ \frac{\left( \overrightarrow{D}^{\left(
i\right) }\left( t\right) e^{-i\omega _it}+h.c.\right) ^2}{n^2\left( \omega
_i\right) }+\left( \overrightarrow{B}^{\left( i\right) }\left( t\right)
e^{-i\omega _it}+h.c.\right) ^2\right] \right. \\  \\ 
\left. -2\left[ \gamma ^{\left( 2\right) }\left( \omega _3=\omega _1+\omega
_2\right) \vdots \overrightarrow{D}^{\left( 3\right) +}\left( t\right) 
\overrightarrow{D}^{\left( 2\right) }\left( t\right) \overrightarrow{D}%
^{\left( 1\right) }\left( t\right) +h.c.\right] \right\} 
\end{array}
\end{equation}
In the derivation, the holo-exchange symmetry of the tensor $\chi ^{\left(
2\right) }$ has been used. In the expansions (7) and (8) of the fields $%
\overrightarrow{D}$ , $\overrightarrow{B}$, only the terms with the
subscripts $\left| \overrightarrow{k}\right| =k_i$, where $k_i=\frac{n\left(
\omega _i\right) \omega _i}c$,$\left( i=1,2,3\right) $, make contributions
to the interaction. So the expansions can be simplified to 
\begin{equation}
\label{38}\overrightarrow{D}^{\left( i\right) }\left( t\right) e^{-i\omega
_it}=i\sqrt{\frac{\hbar \omega _in^2\left( \omega _i\right) }2}\stackunder{%
\left| \overrightarrow{k}\right| =k_i}{\sum }\stackunder{\mu }{\sum }%
\overrightarrow{f}_{\overrightarrow{k}\mu }a_{\overrightarrow{k}\mu }\left(
t\right) , 
\end{equation}

\begin{equation}
\label{39}\overrightarrow{B}^{\left( i\right) }\left( t\right) e^{-i\omega
_it}=c\sqrt{\frac \hbar {2\omega _in^2\left( \omega _i\right) }}\stackunder{%
\left| \overrightarrow{k}\right| =k_i}{\sum }\stackunder{\mu }{\sum }\nabla
\times \overrightarrow{f}_{\overrightarrow{k}\mu }a_{\overrightarrow{k}\mu
}\left( t\right) .
\end{equation}
The function $\overrightarrow{f}_{\overrightarrow{k}\mu }$ can be decomposed
as $\overrightarrow{f}_{\overrightarrow{k}\mu }=f_{\overrightarrow{k}}\left( 
\overrightarrow{r}\right) \cdot $ $\overrightarrow{e}_{\overrightarrow{k}\mu
}$, where $f_{\overrightarrow{k}}\left( \overrightarrow{r}\right) $
satisfies the eigen-equation 
\begin{equation}
\label{40}\nabla ^2f_{\overrightarrow{k}}\left( \overrightarrow{r}\right)
=-\left| \overrightarrow{k}\right| ^2f_{\overrightarrow{k}}\left( 
\overrightarrow{r}\right) 
\end{equation}
Substituting (38) and (39) into Eq.(37), we get the Hamiltonian expressed by
annihilation and creation operators 
\begin{equation}
\label{41}
\begin{array}{c}
\widetilde{H}=\stackunder{\overrightarrow{k}_1,\overrightarrow{k}_2,%
\overrightarrow{k}_3}{\stackrel{\left| \overrightarrow{k}_i\right| =k_i}{%
\sum }}\stackunder{\mu _1,\mu _2,\mu _3}{\sum }\left[ \stackrel{3}{%
\stackunder{i=1}{\sum }}\hbar \omega _ia_{\overrightarrow{k}_i\mu _i}^{+}a_{%
\overrightarrow{k}_i\mu _i}\right.  \\  \\ 
\left. +(\alpha _{\overrightarrow{k}_1\mu _1\overrightarrow{k}_2\mu _2%
\overrightarrow{k}_3\mu _3}\beta _{\overrightarrow{k}_1\overrightarrow{k}_2%
\overrightarrow{k}_3}a_{\overrightarrow{k}_3\mu _3}^{+}a_{\overrightarrow{k}%
_2\mu _2}a_{\overrightarrow{k}_1\mu _1}+h.c.)\right] ,
\end{array}
\end{equation}
where the constant $\beta $ (called phase-matching factor) is defined as 
\begin{equation}
\label{42}\beta _{\overrightarrow{k}_1\overrightarrow{k}_2\overrightarrow{k}%
_3}=\sqrt{V}\int f_{\overrightarrow{k}_3}^{*}f_{\overrightarrow{k}_2}f_{%
\overrightarrow{k}_1}d^3\overrightarrow{r},
\end{equation}
and V is volume of the nonlinear media. The constant $\alpha $ is 
\begin{equation}
\label{43}
\begin{array}{c}
\alpha _{
\overrightarrow{k}_1\mu _1\overrightarrow{k}_2\mu _2\overrightarrow{k}_3\mu
_3}=-i\sqrt{\frac{\hbar ^3\omega _1\omega _2\omega _3}{8Vn^2\left( \omega
_1\right) n^2\left( \omega _2\right) n^2\left( \omega _3\right) }} \\  \\ 
x^{\left( 2\right) }\left( \omega _3=\omega _1+\omega _2\right) \vdots 
\overrightarrow{e}_{\overrightarrow{k}_3\mu _3}^{*}\overrightarrow{e}_{%
\overrightarrow{k}_2\mu _2}\overrightarrow{e}_{\overrightarrow{k}_1\mu _1}
\end{array}
\end{equation}
and it is determined by the polarization matching condition of the optical
field. When $\overrightarrow{k}_1,\overrightarrow{k}_2,\overrightarrow{k}_3$
are collinear, the Hamiltonian (41) can be simplified. Suppose the
polarizations of the optical fields are given (indicated by $\mu
_1^{^{\prime }},\mu _2^{^{\prime }},\mu _3^{^{\prime }}$ respectively) and $%
\overrightarrow{k}_i=k_i\overrightarrow{e}_z$ $\left( i=1,2,3\right) $,
where $\overrightarrow{e}_z$ is the unit vector of z-axis. The expansion
function is approximately a plane wave, i.e., 
\begin{equation}
\label{44}f_{k_i}\left( \overrightarrow{r}\right) =S_{k_i}\left( x,y\right) 
\frac{e^{-ik_iz}}{\sqrt{L}},
\end{equation}
where L is the interaction length. Let $a_i$ represent the operator $%
a_{k_i\mu _i^{^{\prime }}}$ and 
\begin{equation}
\label{45}\alpha =\alpha _{k_1\mu _1^{^{\prime }}k_2\mu _2^{^{\prime
}}k_3\mu _3^{^{\prime }}}\sqrt{\frac VL}\int S_{k_3}^{*}S_{k_2}S_{k_1}dxdy%
\text{,}
\end{equation}
then the Hamiltonian (41) is simplified to 
\begin{equation}
\label{46}\widetilde{H}=\stackrel{3}{\stackunder{i=1}{\sum }}\hbar \omega
_ia_i^{+}a_i+\left( \alpha \cdot \frac{e^{i\Delta kL}-1}{i\Delta kL}%
a_3^{+}a_2a_1+h.c.\right) ,
\end{equation}
where the phase mismatch $\Delta k=k_1+k_2-k_3$. Eq.(46) is often used to
analyze quantum properties of the parametric process [26-28]. Here we give
its exact derivation and determine the expression of the parameter $\alpha $.

The optical field in the third-order nonlinear medium can be quantized in a
similar way. For example, the Hamiltonian of the nondegenerate
four-wave-mixing process [23] with $\omega _3+\omega _4=\omega _1+\omega _2$
is 
\begin{equation}
\label{47}
\begin{array}{c}
\widetilde{H}=\stackunder{\overrightarrow{k}_1,\overrightarrow{k}_2,%
\overrightarrow{k}_3,\overrightarrow{k}_4}{\stackrel{\left| \overrightarrow{k%
}_i\right| =k_i}{\sum }}\stackunder{\mu _1,\mu _2,\mu _3,\mu _4}{\sum }%
\left[ \stackrel{4}{\stackunder{i=1}{\sum }}\hbar \omega _ia_{%
\overrightarrow{k}_i\mu _i}^{+}a_{\overrightarrow{k}_i\mu _i}\right.  \\  \\ 
\left. +(\alpha _{\overrightarrow{k}_1\mu _1\overrightarrow{k}_2\mu _2%
\overrightarrow{k}_3\mu _3\overrightarrow{k}_4\mu _4}\beta _{\overrightarrow{%
k}_1\overrightarrow{k}_2\overrightarrow{k}_3\overrightarrow{k}_4}a_{%
\overrightarrow{k}_3\mu _3}^{+}a_{\overrightarrow{k}_4\mu _4}^{+}a_{%
\overrightarrow{k}_2\mu _2}a_{\overrightarrow{k}_1\mu _1}+h.c.)\right] ,
\end{array}
\end{equation}
where 
\begin{equation}
\label{48}\beta _{\overrightarrow{k}_1\overrightarrow{k}_2\overrightarrow{k}%
_3\overrightarrow{k}_4}=V\int f_{\overrightarrow{k}_4}^{*}f_{\overrightarrow{%
k}_3}^{*}f_{\overrightarrow{k}_2}f_{\overrightarrow{k}_1}d^3\overrightarrow{r%
},
\end{equation}
and 
\begin{equation}
\label{49}
\begin{array}{c}
\alpha _{
\overrightarrow{k}_1\mu _1\overrightarrow{k}_2\mu _2\overrightarrow{k}_3\mu
_3\overrightarrow{k}_4\mu _4}=-i\sqrt{\frac{\hbar ^4\omega _1\omega _2\omega
_3\omega _4}{16V^2n^2\left( \omega _1\right) n^2\left( \omega _2\right)
n^2\left( \omega _3\right) n^2\left( \omega _4\right) }} \\  \\ 
x^{\left( 3\right) }\left( \omega _4=-\omega _3+\omega _2+\omega _1\right) 
\stackrel{\cdot }{\stackunder{.}{:}}\overrightarrow{e}_{\overrightarrow{k}%
_4\mu _4}^{*}\overrightarrow{e}_{\overrightarrow{k}_3\mu _3}^{*}%
\overrightarrow{e}_{\overrightarrow{k}_2\mu _2}\overrightarrow{e}_{%
\overrightarrow{k}_1\mu _1}
\end{array}
\end{equation}
The Hamiltonians (41) and (47) make the foundation for analyzing quantum
properties of the parametric or four-wave-mixing process.\\

{\bf Acknowledgment}

This project was supported by the National Nature Foundation of China.

\end{document}